\documentclass[11pt]{article}
\usepackage{amsmath}
\usepackage{amssymb}
\renewcommand{\title}[1]{%
    \bigskip%
    \begin{center}%
    \Large\bf #1%
    \end{center}%
    \vskip .2in}

\renewcommand{\author}[1]{%
    {\begin{center}
    #1
    \end{center}}}

\begin{document}

\title{\bf{Demystification of  Non-relativistic Theories in Curved Background }}

\bigskip

\centerline{\bf{Rabin Banerjee{\footnote{e-mail: rabin@bose.res.in}}}} 
\bigskip

\centerline{ S. N. Bose National Centre 
for Basic Sciences}
\centerline{ JD Block, Sector III, Salt Lake City, Kolkata -700 106, India}
\bigskip

\centerline{\it{Received honorable mention}}
\bigskip

\centerline{\it{ Essay written for the Gravity Research Foundation 2020 Awards for Essays on Gravitation}}

%\centerline{Submission date: 23 March 2020} 

\vskip 1cm
\begin{abstract}

We discuss a new formalism for constructing a non-relativistic (NR) theory in curved background. Named as galilean gauge theory, it is based on gauging the global galilean symmetry. It provides a systematic algorithm for obtaining the covariant curved space time  generalisation of any NR theory defined in flat space time. The resulting background is just the Newton- Cartan manifold. The example of NR free particle is explicitly demonstrated.
\end{abstract}

\newpage

\section{Introduction }

\smallskip

 The formulation of nonrelativistic (NR) theories in a gravitational background has received considerable attention recently. It has found applications in holography, condensed matter systems, fluid dynamics, cosmology and other realted phenomena \cite{SW, j1, ABPR, B}. This formulation is tricky and markedly different from the relativistic case. The lack of a single non-degenerate metric means that it cannot be done by simply modifying the flat space theory by introducing a coupling with the metric. Indeed, the connection of the NR theory in a curved background with its flat space  counterpart appears shrouded in mystery \cite{SW, ABPR, PPP}.
 Over the last few years we have developed a new method which, apart from other things, demystifies this aspect. Named as galilean gauge theory (GGT), it has been fruitfully applied to a wide range of problems, reproducing existing results and also yielding new ones with fresh insights \cite{BMM1, BMM2, BM4}. Stripped of technicalities GGT answers the question, given a NR theory in a flat background what would be the corresponding theory in a curved background. It is a simple, systematic and compact algorithm that is universally applicable, and is trivially able to recover the original galilean symmetry of the flat model, which has been a very thorny point in such analysis.
  
  In this essay we give a self contained exposition of GGT, particularly since such a presentation is lacking. Indeed, during its development, several technical issues had to be dealt with,  which clouded the physical aspects. In the next section we discuss the basic formalism which is followed by an explicit example, the NR free particle in curved background \cite{BM8}.
  
  \bigskip
  
  \section{Basic Formalism of Galilean Gauge Theory (GGT)}
  
 The action for a NR theory in flat space time is (quasi)invariant under the 
 global galilean transformations,
  \begin{equation}
x^\mu \to x^\mu + \zeta^\mu ; \zeta^0 =  -\epsilon, \zeta^k = \omega^k{}_l x^l+ \epsilon^k - v^k t \label{globalgalilean0} 
\end{equation}
along with appropriate transformations for the fields, where the various parameters associated with translations, rotations and boosts are all constants. 

In order to construct the corresponding action in a curved background we first gauge the galilean symmetry by making the parameters local i.e. space time dependent. Now  $\omega^k{}_l$, the 3-dim rotation parameters, $\epsilon^k $,
parameters for  space translations and $v^k$,  the Galilean boosts, are taken to be space time dependent. Finally, recalling the universal role of time in NR theory, the time translation parameter $\epsilon$ is a function of time only. 

The second step is to appropriately modify the action so that it is invariant under the local galilean transformations. Obviously the original action is not invariant since the derivatives no longer transform as they did earlier for global parameters. The trick is to replace the original ordinary derivatives by `covariant' dervatives,  introducing new fields. 
We now require that the covariant derivatives transform, under local galilean transformations, in the same way as ordinary derivatives did under global galilean transformations. This will also fix the transformations of the new fields. Now if a new action is written replacing the ordinary derivatives in the old action by their covariant versions, it will be invarint under the local galilean transformations. 

The third  step is to abstract a geometrical interpretation from the  modified action.  Observe that 
once the galilean symmetry has been localised, it is necessary to introduce local coordinate bases at every point of space time which are trivially connected with the global coordinates {\footnote{Indices from the beginning of the alphabet ($\alpha, \beta, \,\, a, b, ..$) denote local coordinates while global ones are indicated from the middle of the alphabet ($\mu, \nu,\,\,    i, j,.. $). Greek letters denote space time while nly space is indicated by Latin alphabets}},
\begin{equation}
e^\alpha = \delta^\alpha_\mu e^\mu
\label{basis}
\end{equation}
at this stage. Later this connection will become nontrivial. 

To see how this happens, consider the galilean transformations of derivatives that follow from (\ref{globalgalilean0}),

\begin{equation}
\delta \dfrac{dx^{0}}{d\lambda} = \dfrac{d}{d\lambda}(\delta x^{0})= 0\,\,\,;\,\,\,\delta \dfrac{dx^{k}}{d\lambda} = w^{k}_{j}\dfrac{dx^{j}}{d\lambda} - v^{k}\dfrac{dx^{0}}{d\lambda}
\label{der1}
\end{equation}
%\vspace*{25mm}
as $\delta x^0=-\epsilon $ is constant.
For achieving invariance under local galilean transformations, as already stated, covariant derivatives ($\frac{D}{d\lambda}$) have to be introduced. The simplest possibility is a straightforward extension of (\ref{basis}),

\begin{equation}
\dfrac{Dx^{\alpha}}{d\lambda}  = \dfrac{dx^{\mu}}{d\lambda} \Lambda^{\alpha}_{\mu}
\label{red}
\end{equation}
 
Here $\lambda$ is an arbitrary parameter. For $\Lambda^{\alpha}_{\mu}= \delta^\alpha_\mu$ the covariant derivative reduces to the ordinary derivative, mimicking (\ref{basis}). The meaning of the new variable $\Lambda$ will soon become clear. From our algorithm the transformations of covariant derivatives and ordinary derivatives must be form invariant, so that, they also satisfy relations like (\ref{der1}).
From these relations,  using the local form of the galilean transformations to compute the variations, we obtain the transformation law for the $\Lambda$ variable,
\begin{equation}
\delta\Lambda^a_\nu= - \partial_\nu\zeta^\beta\Lambda_\beta^a + \omega^a_b \Lambda^b_\nu - u^a \Lambda^0_\nu\,\,\, ;\,\,\, \delta\Lambda^0_0=\dot\epsilon\Lambda^0_0 \,\,\, ; \,\,\,\Lambda^0_i=0
\label{geometry}
\end{equation}
where the overdot on $\epsilon$ denotes a derivative with respect to $\lambda$. This shows that a geometric interpretation is feasible for the $\Lambda$ variables. From (\ref{geometry}) we find that while the (local) indices $a$ are Lorentz rotated, the (global) indices $\nu$ are coordinate transformed. 
The local galilean transformation is now interpreted as a NR general coordinate transformation and  $\Lambda^\alpha_\nu$ is regarded as the inverse vielbein connecting the local and global basis in a curved space time. The vielbein $\Sigma$ is defined as the inverse of $\Lambda$. In a nutshell,
\begin{equation}
\hat e_\mu=\Lambda_\mu^\alpha \hat e_\alpha\,\,;\,\, \hat e^\mu=\Sigma^\mu_\alpha \hat e^\alpha\,\,;\,\, \Lambda^\alpha_\mu\Sigma^\mu_\beta=\delta^\alpha_\beta\,\,;\,\, \Lambda_\mu^\alpha\Sigma_\alpha^\nu = \delta^\nu_\mu\,\,\,;\,\,\,\hat e^\mu \hat e_\mu= \hat e^\alpha \hat e_\alpha
\label{nut}
\end{equation}
Thus the gauging of the galilean symmetry has led to a curved space generalisation of (\ref{basis}). Passage to the flat limit from (\ref{nut}) is trivially obtained by replacing $\Sigma$ and $\Lambda$ by Kroneckar deltas, since there is no difference between the local and global bases. 

To further advance the geometrical interpretation, we show that combinations of the vielbeins define the elements of Newton-Cartan NC geometry. These are given by,
\begin{equation}
h^{\mu\nu}={\Sigma_a}^{\mu}{\Sigma_a}^{\nu}\,\,\,; \hspace{.2cm}\tau_{\mu}={\Lambda_\mu}^{0} =\Theta\delta_\mu^0\,\,\,;\,\,\,h_{\nu\rho}=\Lambda_{\nu}{}^{a} \Lambda_{\rho}{}^{a}\,\,\, ; \hspace{.2cm}\tau^{\mu}=
{\Sigma_0}^{\mu}\hspace{.3cm}
\label{spm}
\end{equation}
which satisfy the well known NC algebra,
\begin{equation}
  h^{\mu\nu}  \tau_\nu =  h_{\mu\nu} \tau^\nu =0,\,\,\,  \tau^\mu\tau_\mu=1,\,\,\  h^{\mu\nu} h_{\nu\rho}=
 \delta^\mu_\rho- \tau^\mu \tau_\rho 
 \label{nc2}\end{equation}
 
Now comes the final step that yields the theory in the curved background. The theory  obtained by replacing ordinary derivatives by covariant derivatives will be  invariant under the local galilean transformations. It will involve the NC elements through the  veilbeins. The final theory so obtained is interpreted as the curved (NC) space generalisation of the original flat space theory. 
Let us work out an example.

\section{Non-relativistic free particle}
The  parametrized action for a NR free particle in $3$ dimensional Euclidean space and absolute time
  is given by,
\begin{equation}
S = 
\int
\dfrac{1}{2}m\dfrac{ \dfrac{dx^{a}}{d\lambda}\dfrac{dx^{a}}{d\lambda}}{\bigg(\dfrac{dx^{0}}{d\lambda}\bigg)} ~d\lambda
\label{actionum}
\end{equation}
which is  invariant under the global galilean transformations (\ref{globalgalilean0}). After the gauging process, the new action is obtained by replacing the ordinary derivatives by covariant derivatives. Hence it takes the form,  
\begin{equation}
S = 
\int
\dfrac{1}{2}m\dfrac{ \dfrac{Dx^{a}}{d\lambda}\dfrac{Dx^{a}}{d\lambda}}{\bigg(\dfrac{Dx^{0}}{d\lambda}\bigg)} ~d\lambda
\label{actionum1}
\end{equation}
This action has a nice geometrical interpretation. Using (\ref{red}) and (\ref{spm}) it simplifies,
\begin{equation}
 S=  \left(\dfrac{m}{2}\right)\int h_{\nu\rho}\frac{{x'^\nu} x'^{\rho}}{\Theta x'^0} ~d\lambda
 = \left(\dfrac{m}{2}\right)\int h_{\nu\rho}\frac{{x'^\nu} x'^{\rho}}{\tau_\sigma  x'^\sigma} ~d\lambda
\label{lagm1}
\end{equation}
where a prime denotes an ordinary derivative with respect to $\lambda$. Note that it has been expressed in a covariant form using the NC structures. The above action is therefore  interpreted as the action for a NR particle coupled to a NC background. The equation of motion obtained from here is the geodesic equation with the standard Dautcourt connection. The flat space result (\ref{actionum}) is reproduced by setting $h_{0\mu} =0, h_{ij}=\delta_{ij}, \Theta =1$, following from (\ref{spm}).

\section{Conclusions}

 So far we were confined to particle models where the coordinates are the `field' variables. The same recipe holds for an actual field theory. Then the galilean transformations (\ref{globalgalilean0}) have to be augmented by corresponding transformations on the field variables and their derivatives. Upon localisation, the derivatives of fields will not transform covariantly. Covariant derivatives of fields have to be introduced as shown here which transform formally under the local galilean transformations as ordinary derivatives do under global ones. The new fields inducted by this gauging prescription contain, apart from the  veilbeins, other variables that gauge the (field) rotations and boosts. A modified action is written, replacing the ordinary derivatives by covariant ones, which has a geometric interpretation, akin to the particle model. Thus the method of galilean gauge theory (GGT) is a logical way of construting any NR theory in a Newton Cartan background, starting from a given theory in flat space time.

  \end{document}